\documentclass[aip,jcp,numerical,reprint]{revtex4-1}
\usepackage[version=3]{mhchem} 
\usepackage{color}
\usepackage{graphicx}
\usepackage{threeparttable}
\usepackage{longtable}
\usepackage{enumerate}
\usepackage{amsfonts,amssymb}
\usepackage{amsmath}
\usepackage{stmaryrd}
\usepackage{dcolumn}
\usepackage{bm}
\usepackage{bbm}
\usepackage{algpseudocode}
\usepackage{algorithm}
\usepackage[all,cmtip]{xy}
\usepackage{url}

\bibpunct{}{}{,}{s}{}{\textsuperscript{,}}  

\newcommand{\A}{\mathbf{A}}
\newcommand{\B}{\mathbf{B}}

\newcommand{\M}{\mathbf{M}}
\newcommand{\X}{\mathbf{X}}
\newcommand{\W}{\mathbf{W}}
\newcommand{\Y}{\mathbf{Y}}
\newcommand{\U}{\mathbf{U}}
\newcommand{\V}{\mathbf{V}}
\newcommand{\Z}{\mathbf{Z}}
\newcommand{\z}{\mathbf{z}}
\newcommand{\x}{\mathbf{x}}
\newcommand{\y}{\mathbf{y}}
\newcommand{\N}{\mathbf{N}}
\newcommand{\Id}{\mathbf{I}}
\newcommand{\J}{\mathbf{J}}
\newcommand{\K}{\mathbf{K}}
\newcommand{\G}{\mathbf{G}}
\newcommand{\Q}{\mathbf{Q}}

\newcommand{\Hbf}{\mathbf{H}}
\newcommand{\rU}{\mathrm{U}}
\newcommand{\rSp}{\mathrm{Sp}}
\newcommand{\rUSp}{\mathrm{USp}}
\newcommand{\rR}{\mathrm{R}}
\newcommand{\rI}{\mathrm{I}}

\begin{document}

\title{Structured eigenvalue problems in electronic structure methods
from a unified perspective}
\author{Zhendong Li}\email{zhendongli@bnu.edu.cn}
\affiliation{Key Laboratory of Theoretical and Computational Photochemistry, Ministry of Education, College of Chemistry, Beijing Normal University, Beijing 100875, China}
\date{\today}

\begin{abstract}
In (relativistic) electronic structure methods, the quaternion matrix eigenvalue problem and
the linear response (Bethe-Salpeter) eigenvalue problem
for excitation energies are two frequently encountered structured eigenvalue problems.
While the former problem was thoroughly studied, the later problem in its most general form, namely,
the complex case without assuming the positive definiteness of the electronic Hessian,
is not fully understood. In view of their very similar mathematical structures, we examined
these two problems from a unified point of view. We showed that the identification of Lie group
structures for their eigenvectors provides a framework to
design diagonalization algorithms as well as numerical optimizations techniques on the corresponding manifolds.
By using the same reduction algorithm for the quaternion matrix eigenvalue problem,
we provided a necessary and sufficient condition to characterize the different scenarios,
where the eigenvalues of the original linear response eigenvalue problem are real, purely imaginary, or complex.
The result can be viewed as a natural generalization of the well-known condition for the real matrix case.
\end{abstract}
\maketitle

\section{Introduction}
There are two frequently appeared structured eigenvalue problems in
(relativistic) electronic structure
methods, which can be written into a unified way as
\begin{eqnarray}
\M_s\z&=&\z\,\omega,\quad s=\pm1,\nonumber\\
\M_s&=&\left[\begin{array}{cc}
\A & \B \\
-\B^* & s\A^* \\
\end{array}\right],\quad
\z=\left[\begin{array}{cc}
\x \\
\y \\
\end{array}\right],
\nonumber\\
\A^\dagger&=&\A,\quad \B^T = -s \B,\label{Mdef}
\end{eqnarray}
where $\M_s\in\mathbb{C}^{2n\times2n}$, $\z\in \mathbb{C}^{2n}$,
$\A\in\mathbb{C}^{n\times n}$ is Hermitian, and
$\B\in\mathbb{C}^{n\times n}$ is antisymmetric for $s=+1$ or
symmetric for $s=-1$.

The $s=+1$ case appears in matrix representations
of Hermitian operators, such as the Fock operator
of closed-shell systems, in a Kramers paired basis\cite{dyall2007introduction}.
Another example is the equation-of-motion method\cite{rowe1968equations}
for ionization and electron attachment
from a closed-shell reference, where the excitation operator
$O_n^\dagger$ is expanded in a paired basis $\{a_p^\dagger,a_p\}$,
viz., $O_n^\dagger=\sum_{p}(a_p^\dagger X_p-a_q Y_p)$.
The Hermitian matrix $\M_+$ is usually referred as quaternion matrix\cite{rosch1983time,bunse1989quaternion,saue1999quaternion},
since it can be rewritten as
\begin{eqnarray}
\M_+=\mathbf{I}_2\otimes\A_{\rR}+i\sigma_z\otimes \A_{\rI}
+i\sigma_y\otimes \B_{\rR}+i\sigma_x\otimes \B_{\rI}.
\end{eqnarray}
where $\{\mathbf{I}_2,i\sigma_z,i\sigma_y,i\sigma_x\}$ is isomorphic to
the set of quaternion units $\{1,\breve{i},\breve{j},\breve{k}\}$, where
$\A_{\rR}$ (or $\A_{\rI}$) represents the real (or imaginary) part of $\A$.
The corresponding eigenvalue problem is well-studied,
and several efficient algorithms have been presented\cite{rosch1983time,dongarra1984eigenvalue,bunse1989quaternion,saue1999quaternion,shiozaki2017efficient},
based on the generalization of established algorithms for complex matrices to quaternion
algebra or the use of unitary symplectic transformations.

The $s=-1$ case appears in the linear response problem\cite{olsen1985linear,olsen1988solution,sasagane1993higher,christiansen1998response,
casida1995response,gao2004time,bast2009relativistic,egidi2016direct,liu2018relativistic,komorovsky2019four}
for excitation energies of Hartree-Fock (HF), density functional theory (DFT), multi-configurational self-consistent field (MCSCF), or the Bethe-Salpeter equation (BSE)\cite{salpeter1951relativistic}.
Compared with the $s=1$ case, the linear response eigenvalue problem is more challenging since
$\M_-$ is non-Hermitian. In practice, we are mostly interested in the real eigenvalues,
which correspond to physical excitation energies. Unfortunately, the
condition for the existence of all \emph{real} eigenvalues is only partially understood.
In the nonrelativistic\cite{stratmann1998efficient} and some relativistic cases\cite{liu2018relativistic},
where $\M_-$ becomes real, the eigenvalue problem \eqref{Mdef} is equivalent to
the reduced problem
\begin{eqnarray}
(\A-\B)(\A+\B)(\x+\y)=(\x+\y)\omega^2,\label{AmBApB}
\end{eqnarray}
or
\begin{eqnarray}
(\A+\B)(\A-\B)(\x-\y)=(\x-\y)\omega^2.\label{ApBAmB}
\end{eqnarray}
Thus, the eigenvalues of the original problem are all real if and only if
the eigenvalues of the reduced matrix $(\A-\B)(\A+\B)$ (or its transpose $(\A+\B)(\A-\B)$) are all nonnegative,
i.e., $\omega^2\ge 0$. Besides, the use of Eq. \eqref{AmBApB} or \eqref{ApBAmB} also
reduces the cost for diagonalization compared with that for Eq. \eqref{Mdef}.
If $\M_-$ is complex as in the relativistic case in general,
such reduction is not possible. Assuming the positive definiteness of the so-called
electronic Hessian,
\begin{eqnarray}
\left[\begin{array}{cc}
\A & \B \\
\B^* & \A^* \\
\end{array}\right]\succ0,
\end{eqnarray}
one can show that all eigenvalues of $\M_-$ are real\cite{shao2015properties,shao2016structure,benner2018some}.
However, this condition is only a sufficient condition.
In the real case, this implies $\A-\B\succ0$ and $\A+\B\succ0$.
Another sufficient condition is $\B=0$, in which case
$\M_-$ is block-diagonal and all its eigenvalues are real,
even though there can be negative eigenvalues in $\A$.
The situation, where the electronic Hessian is not positive definite but
$\M_-$ still have all real eigenvalues, is also practically meaningful.
It happens in using an excited state as reference,
a trick that has been commonly used in time-dependent DFT to treat
excited states of systems with a spatially degenerate ground state\cite{seth2005calculation}.
Besides, it also happens in scanning the potential energy curves,
where a curve crossing is encountered between the ground and an excited
state with a different symmetry\cite{li2011spin}.
In such cases, the negative eigenvalues correspond to de-excitations to a lower
energy state.

Due to the similarity in mathematical structures for the $s=1$ and
$s=-1$ cases, in this work we examine these two problems from a unified perspective.
We first identify the Lie group structures for their eigenvectors (see Sec. \ref{sec:liegroup}).
Then, by using the same reduction algorithm for the $s=1$ case (see Sec. \ref{sec:reduction}),
we provide a condition as a generalization of the real case based
on the reduced problems (Eqs. \eqref{AmBApB} and \eqref{ApBAmB})
to characterize the different scenarios, where the eigenvalues of the
complex linear response problem are real, purely imaginary, or complex (see Sec. \ref{sec:condition}).
Some typical examples are provided in Sec. \ref{sec:examples} to illustrate the complexity of
the eigenvalue problem \eqref{Mdef} in the $s=-1$ case.

\section{Lie group structures of the eigensystems}\label{sec:liegroup}
The matrices in Eq. \eqref{Mdef} with $s=1$ and $s=-1$ are closely
related with the skew-Hamiltonian matrix $\W$ and Hamiltonian matrix $\Hbf$
in real field\cite{kressner2005}, respectively,
\begin{eqnarray}
\W&=&\left[\begin{array}{cc}
\W_{11} & \W_{12} \\
\W_{21} & \W_{11}^T \\
\end{array}\right],\; \W_{12}=-\W_{12}^T,\; \W_{21}=-\W_{21}^T,\label{SkewH}\\
\Hbf&=&\left[\begin{array}{cc}
\Hbf_{11} & \Hbf_{12} \\
\Hbf_{21} & -\mathbf{H}_{11}^T \\
\end{array}\right],\; \Hbf_{12}=\Hbf_{12}^T,\; \Hbf_{21}=\Hbf_{21}^T.\label{Ham}
\end{eqnarray}
The identification of the Hamiltonian structure for the linear
response problem was presented in Ref. \onlinecite{list2014identifying}
for the real matrix case. It can be shown that the eigenvalues of $\W$ appear
in pairs $\{\omega,\omega\}$, while the eigenvalues of $\Hbf$ appear in pairs $\{\omega,-\omega\}$\cite{kressner2005}.
The same results also hold for complex matrices. Besides,
the additional relations with complex conjugation in Eq. \eqref{Mdef} compared
with Eqs. \eqref{SkewH} and \eqref{Ham}, viz.,
$\W_{21}=-\W_{12}^*=-\mathbf{B}^*$,
$\W_{11}^T=\W_{11}^*=\mathbf{A}^*$,
$\Hbf_{21}=-\Hbf_{12}^*=-\mathbf{B}^*$,
and
$\Hbf_{11}^T=\Hbf_{11}^*=\mathbf{A}^*$,
lead to further structures on eigenvalues and eigenvectors,
\begin{eqnarray}
\Z_s=
\left[\begin{array}{cc}
\X & -s\Y^* \\
\Y &   \X^* \\
\end{array}\right],\quad
\boldsymbol{\omega}_s=
\left[\begin{array}{cc}
\boldsymbol{\omega} &  \mathbf{0} \\
\mathbf{0} & s\boldsymbol{\omega}^* \\
\end{array}\right].\label{eigenvectors}
\end{eqnarray}
Thus, the symmetry relationships among eigenvalues of Eq. \eqref{Mdef} can be deducted.
For $s=1$, the eigenvalues are real doubly degenerate $\{\omega,\omega=\omega^*\}$,
which is a reflection of the time reversal symmetry.
For $s=-1$, the quadruple of eigenvalues $\{\omega, -\omega, \omega^*, -\omega^*\}$ appears.
If $\omega=\omega^*$ (or $-\omega^*$) is real (purely imaginary),
then the quadruple reduces to the pair $\{\omega,-\omega\}$.
Note that the pair structure \eqref{eigenvectors} always holds for $s=1$,
since $(\X,\Y)$ and $(-\Y^*,\X^*)$ are orthogonal. However, for $s=-1$,
the situation becomes more complicated in the
degenerate case $\omega=-\omega^*$, where the pair structure
of eigenvectors does not necessarily hold. The following discussion in this section only
works for the $s=-1$ case where all the eigenvectors have the pair structure \eqref{eigenvectors},
while the algorithm presented in Sec. \ref{sec:condition} does not have this assumption.

While most of the previous studies focused on the paired structure \eqref{eigenvectors}
for a given matrix $\M_s$, we note that if the set of matrices with
the same form as the eigenvectors $\Z_s$ \eqref{eigenvectors} is considered,
along with the commonly applied normalization conditions,
\begin{eqnarray}
\Z^\dagger_s \N_s\Z_s =\N_s,\quad
\N_s=\left[\begin{array}{cc}
\Id_n & \mathbf{0} \\
\mathbf{0} & s\Id_n
\end{array}\right],\label{normal}
\end{eqnarray}
these matrices actually form matrix groups, viz.,
\begin{eqnarray}
\mathcal{G}_s =\{\Z_s:\,
\Z_s=\left[\begin{array}{cc}
\X & -s\Y^* \\
\Y &   \X^* \\
\end{array}\right],\,\Z^\dagger_s \N_s\Z_s =\N_s\}.\label{Gs}
\end{eqnarray}
This can be simply verified by following the definition of groups as follows:

(1) This set is closed under multiplication of two matrices, since
\begin{eqnarray}
\Z_{s,1}\Z_{s,2}
&=&
\left[\begin{array}{cc}
\X_1 & -s\Y^*_1 \\
\Y_1 &   \X^*_1 \\
\end{array}\right]
\left[\begin{array}{cc}
\X_2 & -s\Y^*_2 \\
\Y_2 &   \X^*_2 \\
\end{array}\right]\nonumber\\
&=&
\left[\begin{array}{cc}
\X_1\X_2-s\Y^*_1\Y_2 & -s(\X_1\Y^*_2+\Y^*_1\X^*_2) \\
\X^*_1\Y_2+\Y_1\X_2 &  \X^*_1\X^*_2-s\Y_1\Y^*_2 \\
\end{array}\right]\nonumber\\
&\triangleq&
\left[\begin{array}{cc}
\mathcal{X} & -s\mathcal{Y}^* \\
\mathcal{Y} &   \mathcal{X}^* \\
\end{array}\right],
\end{eqnarray}
and $(\Z_{s,1}\Z_{s,2})^\dagger\N_s(\Z_{s,1}\Z_{s,2})=\N_s$,
that is $\Z_{s,1}\Z_{s,2} \in \mathcal{G}_s$.

(2) The identity element is just $\Id_{2n}$.

(3) The inverse of any element $\Z_s$ exists, since $\Z_s$ satisfies
the normalization condition \eqref{normal},
\begin{eqnarray}
\Z_s^{-1}
=\N_s\Z_s^\dagger\N_s
=\left[\begin{array}{cc}
\X^\dagger & s\Y^\dagger \\
-\Y^T & \X^T
\end{array}\right]\in \mathcal{G}_s.\label{zinv}
\end{eqnarray}

In fact, the groups \eqref{Gs} are just the
unitary symplectic Lie groups, viz., $\rUSp(2n)=\rU(2n)\cap \rSp(2n,\mathbb{C})$ for $s=+1$,
and $\rUSp(n,n)=\rU(n,n)\cap \rSp(2n,\mathbb{C})$ for $s=-1$.
Here, $\rU(p,q)$ represents the unitary group with signature $(p,q)$, i.e.,
\begin{eqnarray}
\U(p,q)=\{\M:\;\M^\dagger \Id_{p,q}\M = \Id_{p,q}\},\;\;
\Id_{p,q}=\left[
\begin{array}{cc}
\Id_{p} & \mathbf{0} \\
\mathbf{0} & -\Id_{q} \\
\end{array}\right],\label{Upq}
\end{eqnarray}
and $\rSp(2n,\mathbb{C})$ represents the complex symplectic
group,
\begin{eqnarray}
\rSp(2n,\mathbb{C})=\{\M:\;\M^T \J\M = \J\},\;\;
\J=\left[\begin{array}{cc}
\mathbf{0} & \Id_{n} \\
-\Id_{n} & \mathbf{0}\\
\end{array}\right].\label{Sp2n}
\end{eqnarray}
Such equivalence can be established by realizing that
a combination of  the conditions in Eqs. \eqref{Upq} and \eqref{Sp2n} leads to
a condition
\begin{eqnarray}
\J_s\M^*=\M\J_s,\quad
\J_s=\left[\begin{array}{cc}
\mathbf{0} & \Id_{n} \\
-s\Id_{n} & \mathbf{0}\\
\end{array}\right],\label{JMsMJ}
\end{eqnarray}
which implies the block structures in Eq. \eqref{Gs}
for the $2n$-by-$2n$ matrix $\M$. Note in passing that
the case $s=1$ for Eq. \eqref{JMsMJ} reveals the
underlying commutation between an operator
and the time reversal operator.

The identification of Lie group structures for $\mathcal{G}_s$
has important consequences. In particular, it simplifies the
design of structure-preserving algorithms. For instance,
for the case $s=1$, the Lie algebra $\mathfrak{usp}(2n)$
corresponding to the Lie group $\rUSp(2n)$ is
\begin{eqnarray}
\mathfrak{usp}(2n)=\{\K:\;
\K^\dagger =-\K,\;
\J\K^*=\K\J\},\label{usp}
\end{eqnarray}
where the matrix $\K$ can be written more explicitly as
\begin{eqnarray}
\K=\left[\begin{array}{cc}
\K_{11} & -\K_{21}^* \\
\K_{21} & \K_{11}^* \\
\end{array}\right],\;
\K_{11}^\dagger = -\K_{11},\;
\K_{21}^T=\K_{21}.
\end{eqnarray}
This applies to the construction of time reversal adapted
basis operators\cite{aucar1995operator}.
Furthermore, the exponential map $\exp(\K)$ can be used to
transform one Kramers paired basis into another Kramers
paired basis, which was previously used in the Kramers-restricted MCSCF\cite{fleig1997spinor}. More generally,
such Lie group \eqref{Gs} and Lie algebra structures for
the case $s=1$ also implies the possibility to apply the numerical
techniques for the optimization on manifolds\cite{absil2009optimization}
to relativistic spinor optimizations while preserving
the Kramers pair structure. Due to the unified framework presented here,
one can expect that the similar techniques can also be applied to the $s=-1$ case.

\section{Reduction algorithm for the $s=1$ case}\label{sec:reduction}
In this section, we will briefly recapitulate the reduction algorithm for the $s=1$ case,
by adapting the techniques developed for the real skew Hamiltonian matrices\cite{paige1981schur,van1984symplectic}
to the complex matrix $\M_+$ \eqref{Mdef}. Such techniques also form basis for developing diagonalization
algorithms for the relativistic Fock matrix\cite{dongarra1984eigenvalue,shiozaki2017efficient}.
The essential idea is to realize that the unitary symplectic transformation
\begin{eqnarray}
\G=\left[\begin{array}{cc}
\U & -\V^* \\
\V & \U^* \\
\end{array}\right]\in\rUSp(2n),
\end{eqnarray}
when acting on a skew Hamiltonian (complex) matrix, such as
$\W$ \eqref{SkewH} via $\tilde{\W}=\G\W\G^\dagger$, will preserve
the skew-Hamiltonian structure, viz.,
\begin{eqnarray}
\tilde{\W}_{22}=\tilde{\W}_{11}^T,\;\;
\tilde{\W}_{12}^T=-\tilde{\W}_{12},\;\;
\tilde{\W}_{21}^T=-\tilde{\W}_{21}.
\end{eqnarray}
Then, there is a constructive way\cite{paige1981schur} to eliminate the lower-left block of the matrix $\W$ \eqref{SkewH}
by a series of unitary symplectic Householder and Givens transformations,
such that the transformed matrix $\tilde{\W}$ is in the following Paige/Van Loan (PVL) form\cite{kressner2005},
\begin{eqnarray}
\tilde{\W}=\G\W\G^{\dagger}=
\left[\begin{array}{cc}
\tilde{\W}_{11} & \tilde{\W}_{12} \\
\mathbf{0} & \tilde{\W}_{11}^T \\
\end{array}\right].\label{PVLform}
\end{eqnarray}
The crucial point for being able to transform $\W$
into the form \eqref{PVLform} is $\mathbf{W}_{21}=-\mathbf{W}_{21}^T$, and such
property can be preserved during the transformations,
see Ref. \onlinecite{paige1981schur} for details of the transformations.
The computational scaling of such transformation is cubic in the dimension
of the matrix.

Applying this result to $\M_+$ and realizing that the transformed matrix $\tilde{\M}_+$
is still Hermitian, one can conclude that $(\tilde{\M}_+)_{12}=0$ and
$\tilde{\M}_{11}^T=\tilde{\M}_{11}^*$, viz.,
$\tilde{\M}_+$ becomes block-diagonal\cite{shiozaki2017efficient}.
Then, the eigenvalues can be obtained by diagonalizing the Hermitian matrix $\tilde{\M}_{11}$ by
a unitary matrix $\bar{\U}$, which reduces the computational cost compared with the original problem.
The final solution to the original problem can be obtained as
\begin{eqnarray}
\Z_+=\G\left[\begin{array}{cc}
\bar{\U} & \mathbf{0} \\
\mathbf{0} & \bar{\U}^* \\
\end{array}\right]=
\left[\begin{array}{cc}
\U\bar{\U} & -\V^*\bar{\U} \\
\V\bar{\U}^* & \U^*\bar{\U}^* \\
\end{array}\right]\in \mathcal{G}_{+},
\end{eqnarray}
which still preserves the structure \eqref{Gs} due to
the closeness of groups.

\section{Reduction algorithm for the $s=-1$ case}\label{sec:condition}
Due to the non-Hermicity of $\M_-$, a straightforward generalization
of the above reduction algorithm to $\M_-$ using the transformation
in $\rUSp(n,n)$ does not seem to be possible. Because the validity of the
form \eqref{eigenvectors} depends on the properties of the eigensystem.
In some cases, $\M_-$ cannot be diagonalizable, e.g.,
$\M_-=\left[\begin{array}{cc}
1 & 1 \\
-1 & -1 \\
\end{array}\right]$.
To avoid such difficulty, following the observation for
the Hamiltonian matrix by van Loan\cite{van1984symplectic},
one can find the square matrix $\M_-^2$ is a skew
Hamiltonian matrix,
\begin{eqnarray}
\M^2_-&=&
\left[\begin{array}{cc}
\A^2-\B\B^* & \A\B-\B\A^* \\
-\B^*\A+\A^*\B^* & (\A^*)^2-\B^*\B\\
\end{array}\right]\nonumber\\
&\triangleq&
\left[\begin{array}{cc}
\mathcal{A} & \mathcal{B} \\
\mathcal{B}^* & \mathcal{A}^* \\
\end{array}\right],\label{M2}
\end{eqnarray}
where
\begin{eqnarray}
\mathcal{A} &=& \A^2-\B\B^*,\nonumber\\
\mathcal{B} &=& \A\B-\B\A^*,\nonumber\\
\mathcal{A}^\dagger &=& (\A^\dagger)^2-\B^T\B^\dagger
=\A^2-\B\B^* =\mathcal{A},\nonumber\\
\mathcal{B}^T &=& \B^T\A^T-\A^\dagger\B^T
=\B\A^*-\A\B = -\mathcal{B}.
\end{eqnarray}
Thus, it can be brought into the PVL form \eqref{PVLform} using unitary symplectic transformations
as for $\M_+$. However, it deserves to be pointed out that unlike for the Hermitian
matrix $\M_+$, such transformations will not
preserve the relation between the lower-left and upper-right blocks of $\M_-^2$, such that
while $(\tilde{\M}^2_-)_{21}$ is made zero in the transformed matrix, $(\tilde{\M}^2_-)_{12}$ is nonzero.
But the upper-left block $(\tilde{\M}^2_-)_{11}$ can still be used
to compute $\omega^2$ with reduced cost for diagonalization.

This basically gives a criterion for the different scenarios of eigenvalues of the linear response problem.
However, to make better connection to the conditions \eqref{AmBApB} and \eqref{ApBAmB} for the real case.
We use the following transformation\cite{shao2016structure},
\begin{eqnarray}
\Q = \frac{1}{\sqrt{2}}\left[
\begin{array}{cc}
\Id_n & \Id_n \\
i\Id_n & -i\Id_n \\
\end{array}\right]
\end{eqnarray}
which transforms $\M_-$ into a purely imaginary matrix
\begin{eqnarray}
\Q\M_-\Q^\dagger=i\left[
\begin{array}{cc}
(\A+\B)_{\rI} & -(\A-\B)_{\rR} \\
(\A+\B)_{\rR} & (\A-\B)_{\rI} \\
\end{array}\right].
\end{eqnarray}
Then, the square matrix is transformed into a real skew-Hamiltonian matrix,
\begin{widetext}
\begin{eqnarray}
\mathbf{\Pi} \triangleq \Q\M_-^2\Q^\dagger =
\left[\begin{array}{cc}
(\A-\B)_{\rR}(\A+\B)_{\rR}-(\A+\B)_{\rI}(\A+\B)_{\rI} &
(\A-\B)_{\rR}(\A-\B)_{\rI}+(\A+\B)_{\rI}(\A-\B)_{\rR} \\
-(\A+\B)_{\rR}(\A+\B)_{\rI}-(\A-\B)_{\rI}(\A+\B)_{\rR} &
(\A+\B)_{\rR}(\A-\B)_{\rR}-(\A-\B)_{\rI}(\A-\B)_{\rI} \\
\end{array}\right].\label{pi}
\end{eqnarray}
\end{widetext}
It is clear that if $\mathbf{M}_-$ is real, then $\mathbf{\Pi}$ becomes diagonal,
with the diagonal blocks being simply $(\A-\B)(\A+\B)$ and $(\A+\B)(\A-\B)$, and
the transformed eigenvalue problem becomes equivalent to Eqs. \eqref{AmBApB} and \eqref{ApBAmB}, since
\begin{eqnarray}
\mathbf{Q}\z=\mathbf{Q}
\left[\begin{array}{cc}
\x \\
\y \\
\end{array}\right]
=\frac{1}{\sqrt{2}}
\left[\begin{array}{cc}
\x+\y \\
i(\x-\y) \\
\end{array}\right].
\end{eqnarray}
By applying unitary symplectic transformations to reduce $\tilde{\mathbf{\Pi}}=\G\mathbf{\Pi}\G^\dagger$
into the PVL form \eqref{PVLform}, the real non-Hermitian matrix
$\tilde{\mathbf{\Pi}}_{11}$ can be used to compute eigenvalues of $\M_-^2$.
Suppose the eigenvalues of $\tilde{\mathbf{\Pi}}_{11}$ are denoted by $\lambda$, then
we have the following three cases:

(1) $\lambda=\lambda_{\rR}\ge 0$: the pair of real eigenvalues of $\M_-$ is $\{\sqrt{\lambda_{\rR}},-\sqrt{\lambda_{\rR}}\}$.

(2) $\lambda=\lambda_{\rR}<0$: the pair of purely imaginary eigenvalues of $\M_-$ is $\{i\sqrt{-\lambda_{\rR}},
-i\sqrt{-\lambda_{\rR}}\}$.

(3) $\lambda=\lambda_{\rR}+i\lambda_{\rI}$ is complex: $\lambda^*=\lambda_{\rR}-i\lambda_{\rI}$ will also be an eigenvalue of $\tilde{\mathbf{\Pi}}_{11}$, and the quadruple of complex eigenvalues of
$\M_-$ is $\{\omega, -\omega, \omega^*, -\omega^*\}$ with $\omega$ given by
\begin{eqnarray}
\omega = \frac{\zeta}{\sqrt{2}}+i\frac{\lambda_{\rI}}{\sqrt{2}\zeta},\quad
\zeta=\sqrt{\lambda_{\rR}+\sqrt{\lambda_{\rR}^2+\lambda_{\rI}^2}}.
\end{eqnarray}

Thus, the goal to characterize the eigenvalues of the complex linear response problem in a way similar
way as the real case is accomplished based on the eigenvalues of the reduced matrix $\tilde{\mathbf{\Pi}}_{11}$.
In the next section, we will examine some concrete examples.

\section{Illustrative examples}\label{sec:examples}
\begin{figure}\centering
\begin{tabular}{c}
\resizebox{0.4\textwidth}{!}{\includegraphics{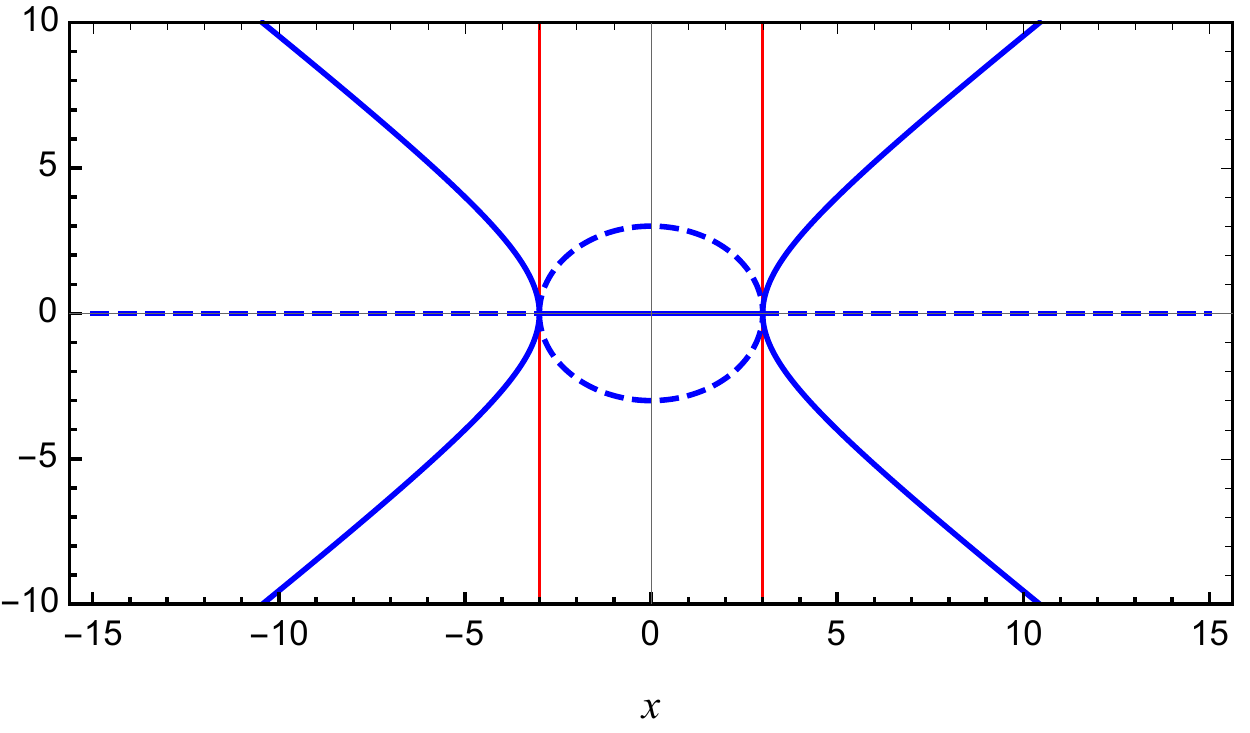}} \\
(a) $\omega$ as a function of $x$ \\
\resizebox{0.4\textwidth}{!}{\includegraphics{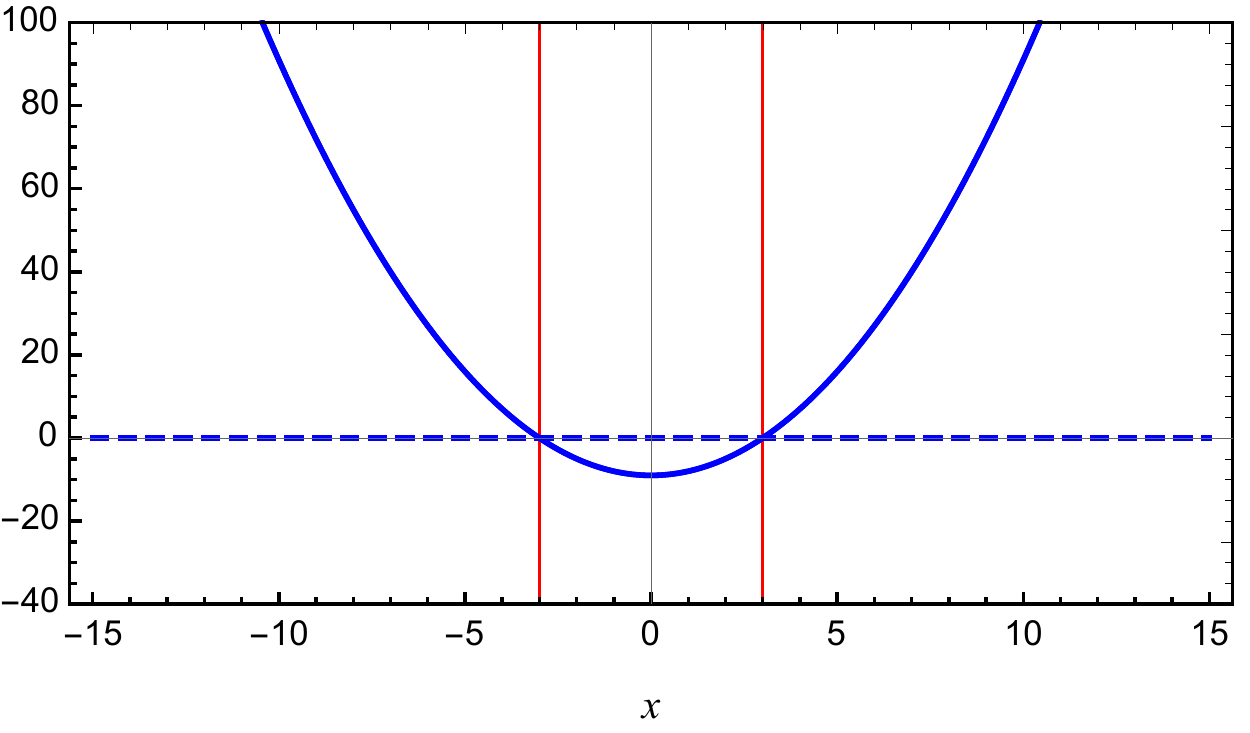}} \\
(b) $\lambda$ as a function of $x$ \\
\end{tabular}
\caption{A 2-by-2 example for $\M_-$ \eqref{example1}:
(a) eigenvalues of $\M_-$ (denoted by $\omega$) as a function of $x$; (b) eigenvalues of $\M_-^2$ (denoted by $\lambda$) as a function of $x$. The blue solid (dashed) lines represent the real (imaginary) parts of eigenvalues.
The two red vertical lines represent critical values of $x=\pm3$.}\label{fig:case1}
\end{figure}

We first illustrate the simplification due to using the square matrix $\M_-^2$ by considering
a 2-by-2 example,
\begin{eqnarray}
\M_-=\left[
\begin{array}{cc}
x & 3i \\
3i & -x \\
\end{array}
\right],\;\;
\M_-^2=\left[
\begin{array}{cc}
x^2-9 & 0 \\
0 & x^2-9 \\
\end{array}
\right],\label{example1}
\end{eqnarray}
where $x\in\mathbb{R}$ is a parameter to mimic the effect of changing physical parameters
in the linear response problem, such as the change of bond length of diatomic molecules in
scanning potential energy curves. It is seen that the matrix $\M_-^2$ is already diagonal,
which gives two identical eigenvalues $\lambda=x^2-9$. Consequently, the original problem
have two eigenvalues $\omega=\pm\sqrt{x^2-9}$. The eigenvalues as a function of $x$ are shown in
Fig. \ref{fig:case1}. The graphs can be classified into three regions:

(1) $x\le-3$: $\lambda\ge 0$ and a pair of real eigenvalues $\omega=\pm\sqrt{x^2-9}$
appears, although the electronic Hessian is not positive definite.

(2) $-3<x<3$: $\lambda<0$ and a pair of purely imaginary eigenvalue $\omega=\pm i\sqrt{9-x^2}$ appear.

(3) $x\ge3$: $\lambda\ge 0$ and the electronic Hessian is positive definite.

\begin{figure}\centering
\begin{tabular}{c}
\resizebox{0.4\textwidth}{!}{\includegraphics{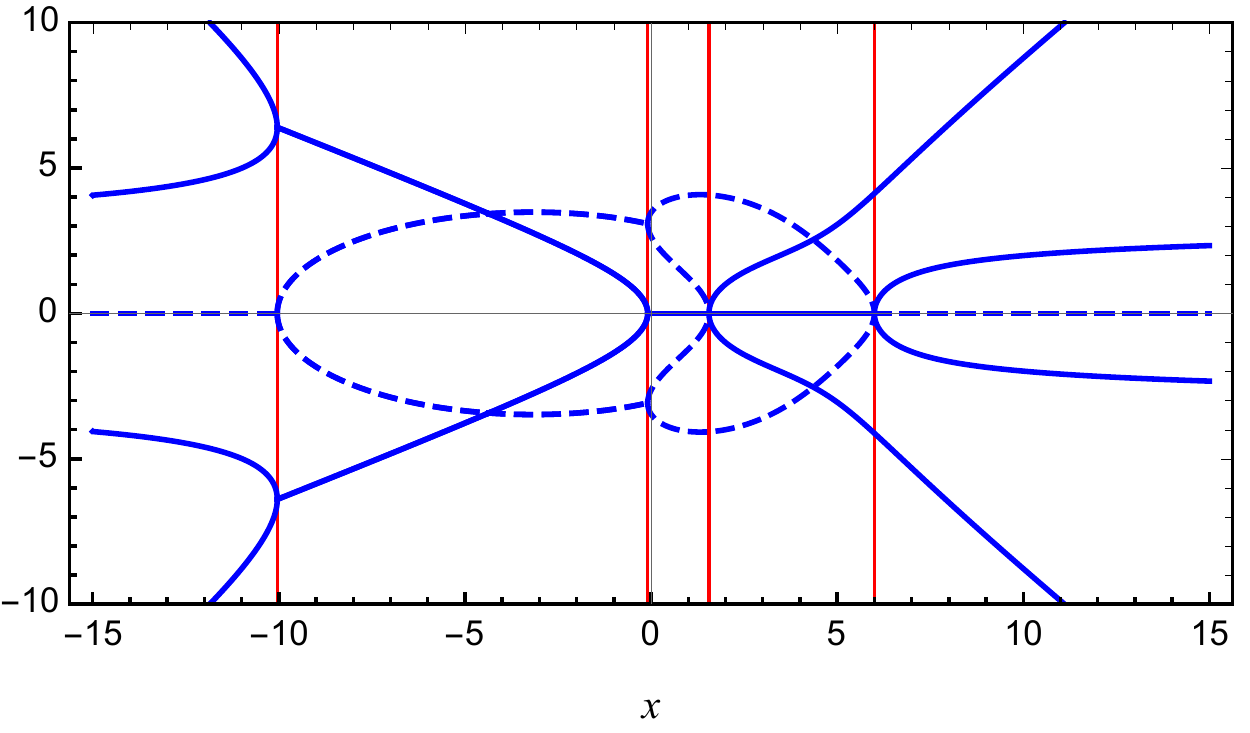}} \\
(a) $\omega$ as a function of $x$ \\
\resizebox{0.4\textwidth}{!}{\includegraphics{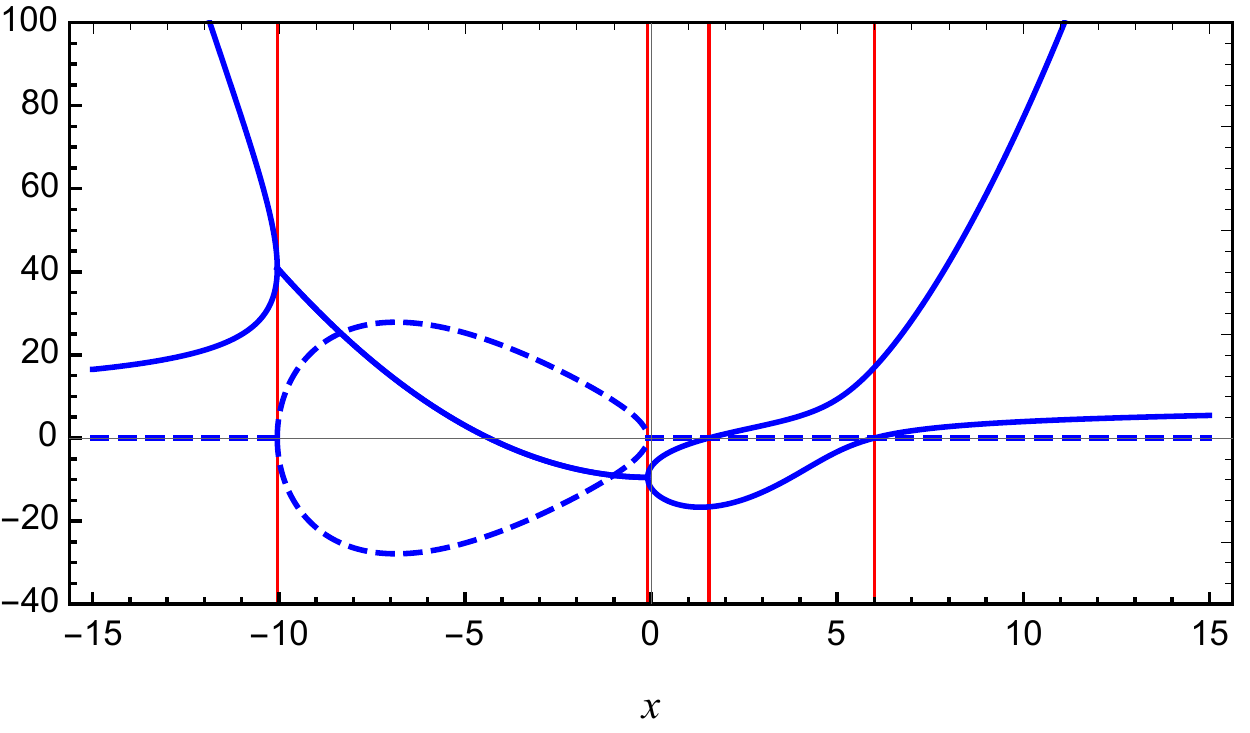}} \\
(b) $\lambda$ as a function of $x$ \\
\end{tabular}
\caption{A 4-by-4 example for $\M_-$ \eqref{example2}:
(a) eigenvalues of $\M_-$ (denoted by $\omega$) as a function of $x$; (b) eigenvalues of $\M_-^2$ (denoted by $\lambda$) as a function of $x$.
The blue solid (dashed) lines represent the real (imaginary) parts of eigenvalues.
The four red vertical lines represent critical values of $x$ given in Eq. \eqref{critical2x}.
}\label{fig:case2}
\end{figure}

Next, we examine a more complex example, which covers all the scenarios for eigenvalues of $\M_-$.
The matrices $\A$ and $\B$ are chosen as
\begin{eqnarray}
\A=\left[
\begin{array}{cc}
x & 3+i \\
3-i & 5 \\
\end{array}
\right],
\quad
\B=
\left[
\begin{array}{cc}
4 & 4 \\
4 & 4 \\
\end{array}
\right].\label{example2}
\end{eqnarray}
The eigenvalues of $\M_-$ can be found analytically as
\begin{eqnarray}
\omega&=&\pm\sqrt{\frac{-19+x^2\pm\sqrt{\Delta}}{2}},\nonumber\\
\Delta&=&25 + 272 x - 74 x^2 + x^4.\label{example2:omega}
\end{eqnarray}
Following the procedure described in the previous section, one can find the corresponding
skew-Hamiltonian $\mathbf{\Pi}$ \eqref{pi} as
\begin{eqnarray}
\mathbf{\Pi}=
\left[\begin{array}{cccc}
-22+x^2 & -37+7x & 0 & -3+x \\
3-x & 3 & 3-x & 0 \\
0 & -13-x & -22+x^2 & 3-x \\
13+x & 0 & -37+7x & 3 \\
\end{array}\right]
\end{eqnarray}
Applying the following Givens rotation $\G$ with an appropriate angle $\theta$ to eliminate $\Pi_{41}$,
\begin{eqnarray}
\G =
\left[\begin{array}{cccc}
1 & 0 & 0 & 0 \\
0 & \cos\theta & 0 & -\sin\theta \\
0 & 0 & 1 & 0 \\
0 & \sin\theta & 0 & \cos\theta \\
\end{array}\right],
\end{eqnarray}
we can find the upper-left block of $\tilde{\mathbf{\Pi}}=\G\mathbf{\Pi}\G^\dagger$ as
\begin{eqnarray}
\tilde{\mathbf{\Pi}}_{11}=
\left[\begin{array}{cc}
-22+x^2 & \frac{\sqrt{2}(-3+x)(-25+3x)}{\sqrt{89+10x+x^2}} \\
-\sqrt{2(89+10x+x^2)} & 3 \\
\end{array}\right].
\end{eqnarray}
It can be verified that its two eigenvalues are given by
$\lambda_\pm=\frac{-19+x^2\pm\sqrt{\Delta}}{2}$, which is
consistent with Eq. \eqref{example2:omega}.
As shown in Fig. \ref{fig:case2}, the eigenvalues $\omega$ and $\lambda$ as a function of $x$ are much more complicated
in this example. The conditions $\Delta=0$ and $\lambda_{\pm}=0$ determine four real critical values of $x$ in total, viz.,
\begin{eqnarray}
&x_1\approx -10.04,\quad x_2\approx -0.09,\nonumber\\
&x_3=\frac{14}{9}\approx 1.56,\quad x_4=6.\label{critical2x}
\end{eqnarray}
Consequently, the graphs can be classified into five regions:

(1) $x\le x_1$: $\lambda_+\ge\lambda_->0$ and $\M_-$ has two pairs of real eigenvalues.

(2) $x_1<x< x_2$: $\lambda_+=\lambda_-^*$ become complex, such that $\M_-$ has a quadruple of eigenvalues.

(3) $x_2\le x< x_3$: $0>\lambda_+\ge\lambda_-$ and $\M_-$ has two pairs of purely imaginary eigenvalues.

(4) $x_3\le x<x_4$: $\lambda_+\ge0>\lambda_-$ and $\M_-$ has a pair of real eigenvalues and a pair of purely imaginary eigenvalues.

(5) $x\ge x_4$: $\lambda_+>\lambda_-\ge0$ and $\M_-$ has two pairs of real eigenvalues.

This example covers all the three different scenarios of eigenvalues of $\M_-$ discussed in the previous section.
All of them can be easily characterized by eigenvalues of a simpler matrix $\tilde{\mathbf{\Pi}}_{11}$ with halved dimension,
which is a natural generalization of $(\A-\B)(\A+\B)$ or $(\A+\B)(\A-\B)$ in the real case.
Finally, we mention that for larger matrices, the eigenvalues cannot be computed analytically, but it is straightforward to
implement the reduction procedure numerically. The behaviors of eigenvalues can be understood in the same way following
the examples presented here.

\section{Conclusion}
In this work, we provided a unified view for the two commonly appeared structured eigenvalue problems
in (relativistic) electronic structure methods - the quaternion matrix eigenvalue problem and
the linear response eigenvalue problem for excitation energies. Using the same
reduction algorithm, we derived a generalized condition
to characterize the different scenarios for eigenvalues of the
complex linear response problem. Such understandings may allow to design
more efficient and robust diagonalization algorithms in future.

\section*{Acknowledgements}
This work was supported by the National Natural Science Foundation of China (Grants
No. 21973003) and the Beijing Normal University Startup Package.

\bibliographystyle{apsrev4-1}
\bibliography{eigenproblem}


\end{document}